# Solutions of Grinberg equation
# and removable cycles in a cycle basis

Heping Jiang

Rm. 702, Building 12, Luomashi Avenue, Xi Cheng district,
Beijing, China, zip code: 100052
E-mail address: jjhpjhp@gmail.com

**Abstract.**

Let $G(V, E)$ be a simple graph with vertex set $V$ and edge set $E$. A generalized cycle is a subgraph such that any vertex degree is even. A simple cycle (briefly in a cycle) is a connected subgraph such that every vertex has degree 2. A basis of the cycle space is called a cycle basis of $G(V, E)$. A cycle basis where the sum of the weights of the cycles is minimal is called a minimum cycle basis of $G$. Grinberg theorem is a necessary condition to have a Hamilton cycle in planar graphs. In this paper, we use the cycles of a cycle basis to replace the faces and obtain an equality of inner faces in Grinberg theorem, called Grinberg equation. We explain why Grinberg theorem can only be a necessary condition of Hamilton graphs and apply the theorem, to be a necessary and sufficient condition, to simple graphs.

## 1. Introduction

Graphs considered in this paper are finite, undirected, and simple graphs. $G(V, E)$ denotes a simple graph with vertex set $V$ and edge set $E$. A Hamilton cycle is a cycle containing every vertex of a graph. A graph is called Hamiltonian if it contains a Hamilton cycle. The Hamilton cycle problem is to find a "good characterization" [1] to be the sufficient and necessary condition of that a graph is Hamiltonian.

$E$ is a family of unordered pairs of elements of $V$. The set $\varepsilon$ of all subsets of $E$ forms an $|E|$-dimensional vector space over $GF_2$ with vector addition $X \oplus Y := (X \cup Y) \setminus (X \cap Y)$ and scalar multiplication $1 \cdot X = X$, $0 \cdot X = \emptyset$ for all X, Y$\in \varepsilon$. A generalized cycle is a subgraph such that any vertex degree is even. A simple cycle (briefly in a cycle) is a connected subgraph such that every vertex has degree 2. All cycles forms a subspace of $(\varepsilon, \oplus, \cdot)$ which are called cycle space of $G$. A basis of the cycle space is called a cycle basis of $G$ $(V, E)$. The dimension of the cycle space is $|V|+|E|-1$. The edges of $G$ have non-negative weights. The weight of a cycle is the sum of the weights of its edges. The weight of a cycle basis is the sum of the weights of its cycles. A cycle basis where the sum of the weights of the cycles is minimum is called a minimum cycle basis of $G$. We use the abbreviation MCB to refer to a minimum cycle basis. A set of cycles is called Hamilton set if the union of these cycles is a Hamilton cycle. A Hamilton set is the result of removing some cycles from a cycle basis of $G$ (if $G$ is a Hamilton graph). There is a mapping of a cycle basis of $G$ onto the graph $G$, which is called surjective mapping. From the mapping point of view, some vertices and edges are the common vertices and the common edges, respectively. We use $R_i$ to refer to an edge that there are $i$ cycles passed through.

In 1968, considering a Hamilton cycle in a planar embedding of the graph $G$ as a closed Jordan curve, E. Grinberg investigated the relations of the outer and inner faces along the curve and received a formula [2], sometimes called Grinberg criterion and later Grinberg theorem, which is a necessary condition to have a Hamilton cycle in planar graph.

**Theorem 1.1.** (Grinberg theorem) Let $G$ be a planar graph with a Hamilton cycle $C$. Then $\sum_{i=1}^{|V|}(i-2)(f_i' - f_i'') = 0$ where $f_i'$ and $f_i''$ are the numbers of faces of degree $i$ contained in inside $C$ and outside $C$, respectively.

Since then, besides using the theorem to investigate the Hamilton graphs in the way of graph order, the study of the condition itself was applied from planar graphs to 2-factor graphs [3], some special grid graphs [4] and the Petersen graphs [5], etc., but no generalized results appeared until now. For what reason that the theorem can only be a necessary condition, no references give an explanation.

Cycle bases of graphs, removable cycles and the inclusion-exclusion principle are basic tools in this paper. Cycle bases used in studying graph theory was initiated by S. MacLane [6] and got more development in algorithm since the cycle bases had a variety of applications in science and engineering [7, 8, 9, 10]. In 1987, J. D. Horton presented the first polynomial time algorithm for finding a minimum cycle basis in a non-negative edge weighted graph [11]. In the proof of the Theorem 1.1, if the faces were replaced with a cycle basis then the formula would not hold. In fact, under the condition of that the given graph is a planar graph consisted of elementary faces (cycles), anyone of these faces (cycles) would be partitioned into two new faces (cycles) when a chord is added to it. But in a cycle of a cycle basis of a graph, adding a chord in a cycle may produce more than two cycles. See Figure 1. Obviously, the result will be uncertainty for Theorem 1.1. So no references have been seen about using basic cycles to replace the faces for Grinberg theorem. In the research of removable cycles, the main consideration concentrates on the cycle removability subjected to the order of a graph [12, 13, 14]. However, because the results of the study on 2-connectibility and Hamiltoncity of a graph are still limited to sufficient conditions [15, 16], it is difficult to determine the Hamiltoncity even the graph remains 2-connected when a cycle is removed. Therefore, no advanced results of applying removable cycles to the Hamilton cycle problem and no references have been seen related to Grinberg theorem.

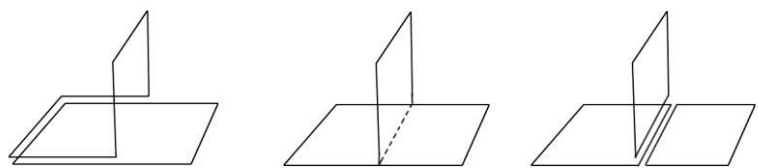

Figure 1

In this paper, instead of the faces of a graph, we use the cycles of a cycle basis to investigate the combinatorial relations of the cycles in a graph. From the combinatorial point of view, by the inclusion-exclusion principle, in the proof of Theorem 1.1, a certain relation between two cycles, that adding a chord in a cycle may produce more than two cycles, was omitted in Hamilton set. Furthermore, in some cases, this omitted item may lead a graph to non-Hamiltonian. In other words, if only there exist such a structure that more than two cycles combined on two common vertices, then there will be some non-Hamiltonian graphs satisfied the Grinberg criterion. It is why Theorem 1.1 can only be a necessary condition of Hamilton graphs. We then present a method to decompose the structure by removing some cycles from a cycle basis so that one can easily determine whether the structure is non-Hamiltonian or not. The main result of this paper improves Theorem 1.1 to be a necessary and sufficient condition for simple graphs.

It is well-known that a graph can be represented by the set $\varepsilon$ and that a Hamilton cycle is the symmetric difference of Hamilton set on a cycle basis. When the faces replaced by basic cycles, we can find, in a mapping of a cycle basis of $G$ onto the graph $G$, a correspondent set of edges and vertices for every cycle in a graph. It is usually that a cycle is represented by its vertex set, that denoting a cycle by its order. Therefore, the combinatorial relation between two cycles will be regard as a relation of two vertex sets. Suppose that the given graph is a Hamilton graph, then every two associated vertex sets in the Hamilton set can be represented by a combination of these sets with two common vertices, See Figure 2, of which union is 2 while others are null. So, by

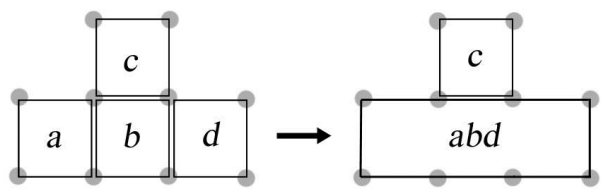

Figure 2

the inclusion-exclusion principle, in Hamilton set, all the vertex sets satisfy the formula for inner faces in Theorem 1.1, that is

**Theorem 1.2.** If $G$ is a Hamilton graph, then $\sum_{i=1}^{|V|}(if'_i - 2f''_i) = |V| - 2$.

We call it Grinberg equation in this paper. While $f'_i$ is called the solution set of the equation, $f''_i$ is called the co-solution set. All solutions of the equation are non-negative. By Theorem 1.2, following corollary is hold,

**Corollary 1.1.** If $G$ has no the solution, then $G$ is not a Hamilton graph.

Theorem 1.2 and Corollary 1.1 imply that we can extend Theorem 1.1 from planar graphs to simple graphs. An analysis of solutions of the equation in Theorem 1.2 we give as the following.

First, there have and only have two kinds of combinatorial relations of two cycles, which union is 2, with two common vertices: combined with common edges or no common edges. In the structure of two cycles combined with common edges there exists a kind of co-solution cycle. If all cycles around this co-solution cycle with common edges are irremovable cycles and this co-solution cycle must be removed from the structure according to the solution, then there will be four $R_1$ edges appeared on the vertex m. See Figure 3. By the rules of constructing a Hamilton cycle, the existence of this kind vertices implies that the structure is non-Hamiltonian. In another word, it is not sufficient for determining Hailtoncity of a graph to have solutions.

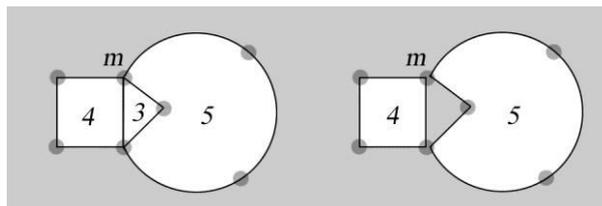

Figure 3

Second, for a given graph a partition based on the vertices may not match with that based on the edges. For example, a trivial graph of order 6 can be partitioned either into two vertex sets of order 3 or two cycles of order 3 and one cycle of order 4. Therefore, from the set point of view, we give the definition of the solution of Grinberg equation, that is a collection of the subset of vertices partitioned by the orders of cycles in the given graph, and which of the union is equal to the order of given graph. This definition implies that even though there exists a solution it may not mean that there have enough cycles to satisfy the solution set in the graph.

From the analysis above, we obtain a new property of Hamilton graphs characterized by the solutions and removable cycles,

**Theorem 1.3.** *G* is a Hamilton graph, if and only if Grinberg equation of *G* has a solution such that the co-solution set is equal to the removable cycle set.

According to theorem 1.3, by deleting the cycles under the following three conditions we can determine whether a graph is Hamiltonian or not (if yes then we derive a Hamilton cycle in the graph). The cycle must be

(1) removable;

(2) a co-solution cycle;

(3) a cycle such that deleting it do not change the removability of other co-solution cycles.

In the next content, section 2 is the proof of Theorem 1.2. In section 3, we give characterizations of the structure that there exists a co-solution cycle, in which every pair of cycles combined with no common edges. Section 4 is the proof of Theorem 1.3. In section 5, we give the brief discussions about how to apply Theorem 1.3 to arbitrary graphs, and whether or not Theorem 1.3 can be a good characterization of Hamilton graphs, particularly applying to a polynomial algorithm.

Terminology and notations not mentioned in this section will be given in the following sections, respectively. Others see [17, 18, 19].

## 2. The proof of theorem 1.2

*Proof.* Let $B$ be a cycle basis of graph $G$, $f_i$ denote a cycle of order $i$, $|f_i|$ refer to the number of $f_i$. $|f| = |f_3|+|f_4|+...+|f_i|$, $i \leq |V|$. Since the union of the whole cycles of $B$ is the given graph G, then we have $|V|=|V_3 \cup V_4 \cup ... \cup V_i|$. By the inclusion-exclusion principle, we have

$$|V| = \sum_{a=3}^{i}|V_a| - \sum_{3 \leq a < b \leq i}|V_a \cap V_b|$$
$$+ \sum_{3 \leq a < b < c \leq i}|V_a \cap V_b \cap V_c| - \cdots +(-1)^{i-1}|V_3 \cap V_4 \cap V_5 \cdots \cap V_i|. \quad (2.1)$$

Suppose $G$ is Hamilton graph, then there exists a Hamilton set and the union of each pair of disjointed cycles is null. Let $|V_a \cap V_b|$ denote the number of the item in the equality (2.1) whose union of each pair of jointed cycles is 2. We then have

$$|V| = \sum_{a=3}^{i}|V_a| - \sum_{3 \leq a < b \leq i}|V_a \cap V_b|. \quad (2.2)$$

Since each pair of jointed cycles combined with a common edge, the edge is an $R_2$ edge. Therefore, the number of the pairs of jointed cycles is equal to the number of $R_2$ edges, that is $|f|-1$. Note that $|V_a \cap V_b|=2$ in the Hamilton set, so $\sum_{3 \leq a < b \leq i}|V_a \cap V_b| = 2(|f|-1)$. Replaced $|f|$ by $|f_3|+|f_4|+...+|f_i|$, $i \leq |V|$, we derive

$$\sum_{3 \leq a < b \leq i}|V_a \cap V_b| = 2(|f_3| + |f_4| + |f_5| + \cdots +|f_i| - 1). \quad (2.3)$$

$\sum_{a=3}^{i}|V_a|$ is the sum of all the subset of vertices partitioned by the orders of cycles, that is

$$\sum_{a=3}^{i}|V_a| = |V_3| + |V_4| + \cdots +|V_i|. \quad (2.4)$$

It is known that $|V_3| = 3|f_3|$, $|V_4| = 4|f_4|$, $\cdots$, $|V_i| = i|f_i|$, so we can write (2.4) as following,

$$\sum_{a=3}^{i}|V_a| = 3|f_3| + 4|f_4| + \cdots +i|f_i|. \quad (2.5)$$

Using (2.3), (2.5) to replace the items in (2.2) respectively, we have

$$\sum_{a=3}^{i} i|f_i| - 2\left(\sum_{a=3}^{i}|f_i| - 1\right) = |V|. \tag{2.6}$$

Note that $|V| = |C|$, and if we denote the inner face $f_i$ by $f_i''$, then (2.6) can be written as

$$\sum_{a=3}^{i}(if_i' - 2f_i') = |C| - 2. \tag{2.7}$$

(2.7) is the equality of inner faces in Theorem 1.1 (Grinberg theorem). We call equality (2.7) Grinberg equation in this paper. Thus, theorem 1.3 holds. □

## 3. The irremovable co-solution cycle

We first give the three rules of constructing a Hamilton cycle and some corollaries.

Rule 1. If a vertex $x$ has degree 2, both edges incident to $x$ must be part of any Hamilton cycle.

Rule 2. No proper subcycle (that is, a cycle not containing all vertices) can be formed when building a Hamilton cycle.

Rule 3. Once the Hamilton cycle is required to use two edges at a vertex $x$, all other edges incident to $x$ must be removed from consideration.

A cycle is removable if we derive a subgraph $G'(V', E')$ such that $V' = V$, $E' = E-1$ by deleting it from a cycle basis of $G(V, E)$ that satisfied the three rules. We use $|\overline{C}_{R1}|$ to denote the number of irremovable cycles such that each cycle has $R_1$ edges and common vertices or common edges, then the following proposition holds.

**Proposition 3.1.** $G$ is non-Hamiltonian if $|\overline{C}_{R1}| \geq 3$.

*Proof.* We assume $G$ is a Hamilton graph. Without losing generality, we need only to consider the case of $|\overline{C}_{R1}|=3$. There have six cases

satisfied the combination of three or more cycles under the given condition. Obviously, there exists at least one vertex in each case that there must have three or more edges so that a Hamilton cycle can pass through, that is contradictory to the assumption. □

Similarly to the next proposition,

**Proposition 3.2.** *G* is non-Hamiltonian if there is a vertex such that $|R_1| \geq 3$.

*Proof* (Omitted)

Let *F* denote a graph, without $|\overline{C}_{R1}| \geq 3$, $|R_1| \geq 3$, of which the equation having solutions. From Theorem 1.2, Corollary 1.1, Proposition 3.1 and 3.2, there have Hamilton graphs in *F* only. So the graphs in the next we considered are the graphs of *F*.

In the proof of Theorem 1.2, since we suppose that the equation of the given graph has a solution, then there have and only have two cases for the combination of cycles whose cardinality of the union is 2, combining with a common edge or without common edges. While in Theorem 1.1 only the case of combining with a common edge was considered. But, in the case of combining without common edges, it may be a non-Hamilton structure.

Let $C_k$ be an $R_1$-free cycle of the set with no removable cycles in *F*. A vertex is called interior if all the jointed edges are $R_2$ and called boundary if there have only two of edges are $R_1$. We have

**Lemma 3.1.** *F* is non-Hamiltonian if *F* has a unique solution such that $C_k$ is a co-solution cycle.

*Proof.* It is clear that all the vertices in Hamilton set are the boundary vertices and $C_k$ is an $R_1$-free cycle in the set of *F*. therefore, if all vertices of $C_k$ are the boundary, then $C_k$ must be a cycle that all its edges are $R_1$. Thus, $C_k$ must not be a co-solution cycle. If $C_k$ is a co-solution cycle, then there must exist at least a non-boundary vertex on the cycle. And, $F$ is a graph without $|\overline{C}_{R1}| \geq 3$ and $|R_1| \geq 3$. So the non-boundary vertices can only be interior. Furthermore, since there are no removable cycles in $F$ and $F$ has a unique solution, then the only way to lead the non-boundary vertices to the interior vertices is to delete $C_k$. And then, all the $R_2$ edges can be changed into $R_1$ edges. By Proposition 3.2, $F$ is non-Hamiltonian. □

## 4. The proof of Theorem 1.3

*Proof.*   For a given $F$, according to the definition of the solution of Grinberg equation, if there are not enough removable cycles in $F$, then $F$ has no Hamilton sets. By Lemma 3.1, there is no Hamilton set in $F$ if $C_k$ is a co-solution cycle. Therefore, in the cases of the combination of cycles whose cardinality of the union is 2, the only case left is that the combining with a common edge. Hence, the given $F$ is a Hamilton set. In addition, suppose that there still have removable cycles in $F$, there must have non-boundary vertices, and then there will be the case that the union of each pair of disjointed cycles is not null. By Theorem 1.2, Grinberg equation does not hold. This is contradictory to the assumption. Therefore, the deleted cycles that satisfied the solution are equal to the removable cycles in $F$ exactly, which implied that the co-solution set is equal to the removable cycle set. □

## 5. Discussion

From the cycle basis points of view, both cycles of order 1 and 2 can be considered inspect of that the graphs considered in this paper are simple graphs. It is noted that the domain of cycles of order 1 is zero and the item of cycles of order 2 in solution is null. So Theorem 1.3 can be applied to arbitrary graphs.

Based on Theorem 1.3, the algorithm consists of three main parts. The first is to find an MCB in a graph; the second is to find the solutions of the equation of the graph, and the third is to determine a cycle satisfied the conditions described in section 1. It is clear that the first two parts have polynomial algorithms [11, 20, 21]. In the third part, we must take three steps. The first step is, under the rules of constructing a Hamilton cycle, to determine whether there have the vertices that $|R_1| \geq 3$ or not in the induced subgraph so that the selected cycle is removable. It means that every vertex selected from the vertex set of the graph should be checked for which one is incident with three or more edges. The second step is to determine that the selected cycle from the first step is a co-solution cycle. The third step is to check the cycle selected from two steps before whether it selected implies the removability of the other co-solution cycles changed. For this aim, we will repeat the first step for the other co-solution cycles. It is easily to estimate that each of the three steps can be solved in polynomial time. Thus, there may have a polynomial algorithm for finding Hamilton graphs.